\documentclass[12pt]{article}

\usepackage{graphicx}
\usepackage{feynmp}
\usepackage{amsmath,amssymb}
\usepackage{bm}
\usepackage{graphicx, color}
\begin{document}
\title{
\begin{flushright}
\ \\*[-80pt] 
\begin{minipage}{0.2\linewidth}
\normalsize
\end{minipage}
\end{flushright}
{\Large \bf 
Renormalizable Quantum Gravity in Low Energy 
without Violating Unitarity
\\*[10pt]}}

\author{
\centerline{
Hajime~Isimori$^{1}$   }
\\*[10pt]
\centerline{
\begin{minipage}{\linewidth}
\begin{center}
$^1${\it \normalsize
Graduate~School~of~Science~and~Technology,~Niigata~University,  
Niigata~950-2181,~Japan} \\ 
\end{center}
\end{minipage}}
\\*[10pt]}

\date{
\centerline{\small \bf Abstract}
\begin{minipage}{0.9\linewidth}
\medskip 
\medskip 
\small
We introduce new techniques that can preserve 
unitarity of the system including ghost particles. 
Negative norms of the particles can be involved in zero-norm 
states by constraints of the physical space. 
These are useful to apply the higher-derivative propagator for quantum gravity 
to suppress divergences of vacuum energy and graviton mass correction. 
The quantum effects are mainly depending on the ghost mass scale. 
As the scale can be chosen in any order, the observed cosmological 
constant is realized. Further, applying ghost partners for the standard model 
particles, quantum gravity with matter fields becomes renormalizable 
with power counting arguments.
\end{minipage}
}

\maketitle

\section{Introduction}

There is a long way to unify 
general relativity and quantum mechanics as in the 
framework of quantum gravity \cite{Carlip,Ashtekar,Kiefer}. 
The two leading candidates of quantum gravity are 
string theory and loop quantum gravity. 
Unfortunately, as any theory of quantum gravity has some deep problem \cite{Rovelli}, 
the theory is not complete yet. 
In general, it is usually thought that 
something new must happen at the Planck scale 
to make a consistent theory of quantum gravity. 
Nevertheless, the scale is too high to reach with current experiments, 
hence, there is no experimental hint for the statement. 
Meanwhile, considering the cutoff scheme, 
observed cosmological constant suggests that the cutoff scale of vacuum energy 
should be around the neutrino mass or micrometer scale. 
To accommodate with the observation, new method that can have 
small energy scale seems necessary.

Perturbative approach of quantum gravity is 
appropriate for low energy \cite{Dewitt,Veltman,Bohr}. Taking the flat-space background, 
one can quantize the weak gravitational field. 
In this method, there appear bad divergences 
in many Feynman diagrams because the coupling constant has negative dimension. 
For instance, quantum correction to the graviton mass squared has 
quartic divergence. If the cutoff scale is the Planck mass, 
graviton mass is correspondent to the Planck mass. 
This is similar to the Higgs mass problem 
though it is renormalizable and less problematic for physics. 
The problem of graviton mass is more serious 
because observations of galaxy and clusters ensure a very long-range force 
but there is no natural way to protect the mass from quantum correction.

Quartic divergence is the worst in the perturbative quantum gravity for one-loop. 
This is the same to the divergence of vacuum energy. 
Then, if some method can solve the cosmological constant 
problem, the problem of perturbative quantum gravity may be also solved. 
The simplest way is to take into account a ghost particle \cite{Moffat}. 
For the sake of opposite sign of commutation relation,  
divergences induced by normal particle (meaning it has positive norm) can be canceled. 
A difficult problem relating to the ghost is the violation of unitarity 
that indicates the violation of probability conservation due to negative norm. 
Although there are several discussions 
and possible solutions e.g. \cite{Hawking,Mannheim}, 
the problem still remains in general understanding. 
This paper suggests new solution to avoid unitarity violation 
by selecting proper physical state relating between normal and ghost particles. 
In particular, the mixing state of them has zero norm 
that plays a key role to hold positive semi-definite norm. 
All the conditions of unitarity can be satisfied by 
the constraints to be physical states.

We propose the existence of ghost partner for graviton. 
To be consistent with large-scale gravity, 
the ghost should have a mass. 
In order to realize the higher-derivative propagator, 
usual graviton is ought to be massive because there is a  
mismatch between the propagators of massless and massive gravitons. 
A mass term of graviton induces negative norm in general 
so that we choose the Pauli-Fierz mass term. 
Taking the term for both gravitons, 
all divergences can be improved 
and the mass squared correction of graviton will be better. 
However, to achieve consistent theory without fine-tuning, 
additional gravitons and their partners are necessary to increase the power of propagator. 
They are assumed to have the same 
quantum number of the original graviton but with different masses. 
If some conditions are satisfied, the theory can be super-renormalizable or completely finite. 
This can be seen as a modified 
version of the Lee-Wick model \cite{Grinstein}.

When we take enough number of gravitons and some conditions, 
quantum effects including vacuum energy are sufficiently suppressed. 
Amounts of the effects are mainly dependent on masses of gravitons 
instead of the cutoff scale. 
As the unitarity violation can be avoided, 
scales of the masses are not need to be 
very high such as the Planck mass. 
Hence we possess the correct value of the cosmological constant 
by taking the graviton masses around the neutrino mass scale. 
This implies that gravity is modified at micrometer scale. 
For the Yukawa-type modification, the inverse square law of gravity 
is confirmed on the scale longer than $56\mu$m at $95\%$ precision \cite{Kapner} 
and more stringent bound about $20\mu$m is recently given by \cite{Rajalakshmi}. 
It is not easy to be consistent with these experiments. 
On the other hand, the loop calculation yields small graviton mass 
which is not negligible even for the case getting the correct vacuum energy. 
Writing the cosmological constant $\lambda$ and the gravitational constant $G$, 
the lightest graviton mass becomes about $\sqrt{G\lambda}$ which 
is of order $10^{-10}\text{pc}^{-1}$. 
This is also close to the experimental bound of the lightest graviton mass \cite{Goldhaber}. 
Thus, the theory of quantum gravity without fine-tuning 
makes modifications for 
short and long distances simultaneously and both are near to current experimental bounds. 
In the sense of possible experimental test, 
quantum gravity with multi-gravitons in low energy is interesting.

The paper is organized as follows. 
In section 2, we explore the model of multi-gravitons with Pauli-Fierz mass terms, 
afterward, the gravitational wave for each graviton is canonically quantized. 
Section 3 introduces the interaction with fermions using connection and covariant 
derivative to calculate the gravitational potential. 
Section 4 explains how to preserve unitarity from ghost particles of 
negative norm by selecting physical state and wave packets. 
In section 5, vacuum energy is calculated with higher-derivative propagator 
mentioning the modification of gravity for short distance 
to keep it viable with experiments of gravity and the cosmological constant. 
In section 6, loop correction of lightest graviton mass is calculated 
with one contribution of a Feynman diagram. 
At last, we have a short conclusion at section 7. 

In our notation, we use $\eta^{\mu\nu}=\textrm{diag}(-1,1,1,1)$ and 
natural units $c=\hbar=1$.

\section{Quantization of gravitons and ghost gravitons}

Throughout this paper, we assume gravitational wave is quantized \cite{Lovas}
which can be done in the same way of other fundamental fields, especially similar 
to photon. Though this is not a standard method to quantize 
gravity, the scheme is convenient to apply higher-derivative propagator 
and to see unitarity of the system. 
Some problems may arise since the quantization 
violates Lorentz invariance and leads unusual interaction terms 
for momentum operator. Further, we need to cancel out many terms inside the Lagrangian 
to quantize the gravitons. 
It looks like very schematic but can be a powerful method. 
As an effective field theory, the system will be valid.

\subsection{Lagrangian}
Let us start with the Einstein-Hilbelt action
\begin{eqnarray}
{\cal L}
=\frac{1}{16\pi G}R+{\cal L}_\text{mat},
\end{eqnarray}
which leads the Einstein equation 
$G_{\mu\nu}=8\pi G T_{\mu\nu}$. 
In the Minkowski space background, the metric 
can be expanded as
\begin{eqnarray}
g_{\mu\nu}
=\eta_{\mu\nu}+\kappa h_{\mu\nu},
\quad
g^{\mu\nu}
=\eta^{\mu\nu}
-\kappa h^{\mu\nu}
+\kappa^2 h^{\mu}{}_{\rho}h^{\rho\nu}
-\cdots.
\end{eqnarray}
Similarly, $\sqrt{-g}$ can be written as 
\begin{eqnarray}
\sqrt{-g}
=1+\frac12\kappa h
-\frac14\kappa^2h^{\mu}{}_{\nu}h^{\nu}{}_{\mu}
+\frac18\kappa^2 h h +\cdots,
\end{eqnarray}
where $h =\eta_{\mu\nu} h^{\mu\nu}$. 
To make the higher-derivative propagator, we explore multi-gravitons model. 
The gravitons are denoted by $h_{\mu\nu}^{(n)}$, 
where $n$ runs from zero to ${\cal N}-1$. 
These fields are dealt with a part of the real gravitational field 
i.e. $h_{\mu\nu}=\sum_n h_{\mu\nu}^{(n)}$. 
All the gravitons 
are assumed to be massive with Pauli-Fierz mass terms
\begin{eqnarray}
{\cal L}_\text{mas}
=-\frac{\kappa^2}{32\pi G\sqrt{-g}}\sum_{n=0}^{{\cal N}-1}
(-1)^{n}m_n^2(h^{(n)}_{\mu\nu}h^{(n)\mu\nu}-h^{(n)}h^{(n)})
\end{eqnarray}
Mass terms of ghost gravitons take opposite signs, 
and relating to those minus signs, ghost kinetic terms 
are modified by
\begin{eqnarray}
\begin{split}
{\cal L}_\text{gho}
=&-
\frac{\kappa^2}{16\pi G\sqrt{-g}}
\sum_{n=0}^{{\cal N}/2-1} 
(h^{(2n+1)\mu\nu}\Box h_{\mu\nu}^{(2n+1)}
-2h^{(2n+1)}_{\mu\nu}\partial^\mu\partial_\sigma h^{(2n+1)\sigma\nu}
\\
&\quad+2h^{(2n+1)}_{\mu\nu}\partial^\mu\partial^\nu h^{(2n+1)}
-h^{(2n+1)}\Box h^{(2n+1)})),
\end{split}
\end{eqnarray}
where we ignore total derivatives. 
Still, it is not enough to make canonical quantization for 
each field $h_{\mu\nu}^{(n)}$. 
To remain only necessary terms, following terms are taken for the cancellation: 
\begin{eqnarray}
\begin{split}
{\cal L}_\text{kin}
=&-\frac{1}{\sqrt{-g}}
\frac{\kappa^2}{32\pi G}
\sum_{m,n,m\not=n} 
(h^{(m)\mu\nu}\Box h_{\mu\nu}^{(n)}
-2h^{(m)\mu\nu}\partial^\mu\partial_\sigma h^{(n)\sigma\nu}
\\
&\quad+2h^{(m)\mu\nu}\partial^\mu\partial_\nu h^{(n)}
-h^{(m)}\Box h^{(n)})).
\end{split}
\end{eqnarray}
The energy momentum tensor is given by 
$T_{\mu\nu}=\frac{2}{\sqrt{-g}}
(\partial_\rho\frac{\partial({\cal L}_\text{mat}\sqrt{-g})}{\partial(\partial_\rho g^{\mu\nu})}
-\frac{\partial({\cal L}_\text{mat}\sqrt{-g})}{\partial g^{\mu\nu}})$, then 
\begin{eqnarray}
\begin{split}
(-1)^n8\pi G T_{\mu\nu}
=&\frac12\kappa
(\partial^{\rho}\partial_\mu h_{\nu\rho}^{(n)}
+\partial^{\rho}\partial_\nu h_{\mu\rho}{}^{(n)}
-\Box h_{\mu\nu}^{(n)}
-\partial_\mu\partial_\nu h^{(n)})
\\
&-\frac12\kappa
(\partial_{\rho}\partial_\sigma h^{(n)\rho\sigma}
-\Box h^{(n)})\eta_{\mu\nu}
+\frac12\kappa m_0^2(h_{\mu\nu}^{(n)}-h^{(n)}\eta_{\mu\nu}).
\end{split}
\end{eqnarray}
Multiplying $\partial^\mu$ for both sides and applying 
$\partial^\mu T_{\mu\nu}=0$ as the first order approximation, 
we get $\partial^\mu h_{\mu\nu}^{(n)}-\partial_{\nu} h^{(n)}=0$. 
Plugging them into the equations above, we get
\begin{eqnarray}
\label{eom}
(\Box-m_0^2) h^{(n)}_{\mu\nu}
=-(-1)^n\frac{16\pi G}{\kappa} 
(T_{\mu\nu}+\frac13(\frac{\partial_\mu\partial_\nu}{m_0^2}-\eta_{\mu\nu})T).
\end{eqnarray}
They can be related to the equations of motion of higher-derivative theory. 
Considering ${\cal N}=2$, they yield 
\begin{eqnarray}
(\Box-m_0^2)(\Box-m_1^2)h_{\mu\nu}
=(m_1^2-m_0^2)
\frac{16\pi G}{\kappa} 
(T_{\mu\nu}+\frac13(\frac{\partial_\mu\partial_\nu}{m_0^2}-\eta_{\mu\nu})T).
\end{eqnarray}

\subsection{Quantization} 
Hereafter, we choose the transverse-traceless gauge (TT-gauge) 
for all gravitational fields. 
If gravitational wave of each field can be quantized we can write
\begin{eqnarray}
 h_{\mu\nu}^{(n)}(x)
=\int \frac{d^3p}{(2\pi)^3}
\frac{1}{\sqrt{2E_{\bm p}^{(n)}}}
\sum_{\lambda}e_{\mu\nu}^{(\lambda)}
(a_{\bm p}^{(\lambda,n)}e^{(-1)^{n}ip\cdot x}
+a_{\bm p}^{(\lambda,n)\dagger}e^{-(-1)^{n}ip\cdot x}),
\end{eqnarray}
where $E_{\bm p}^{(n)}=\sqrt{\bm p^2+m_n^2}$. 
Creation and annihilation operators obey
\begin{eqnarray}
a^{(\lambda,n)}_{\bm p}|0\rangle
=0,
\quad
[a^{(\lambda,n)}_{\bm p},{a^{(\lambda',n')}_{\bm p'}}^\dagger]
=(-1)^{n}\delta_{nn'}\delta_{\lambda\lambda'}\delta^3(\bm p-\bm p').
\end{eqnarray}
Polarization tensors have following properties \cite{Aubert}:
\begin{eqnarray}
e^{(\lambda)}_{\mu\nu}
=e^{(\lambda)}_{\nu\mu},
\quad
e^{(\lambda)\mu}{}_\mu=0,
\quad
p_\mu e^{(\lambda)\mu\nu}=0,
\quad
e_{\mu\nu}^{(\lambda)}
e^{(\lambda')\mu\nu}
=\delta^{\lambda\lambda'}.
\end{eqnarray}
In the TT-gauge, the Einstein tensor with second order of expansion 
can be calculated as
\begin{eqnarray}
G_{\mu\nu}^{(2)}
=-\kappa^2
(\frac12 h_{\rho\sigma}
\partial_\mu\partial_\nu  h^{\rho\sigma}
+ h_{\mu\rho}^{}\Box  h^{\rho}{}_{\nu}
+\frac14  h_{\rho\sigma}\Box  h^{\rho\sigma}\eta_{\mu\nu}).
\end{eqnarray}
Note, the last two terms are automatically canceled if graviton is massless. 
To perform the canonical commutation relations for 
massive gravitons, they are modified by taking
\begin{eqnarray}
{\cal L}_\text{can}
=\frac{\kappa^3}{16\pi G\sqrt{-g}}  
(\sum_{m,n,m\not=n}\frac12 h^{\mu\nu}h_{\rho\sigma}^{(m)}
\partial_\mu\partial_\nu h^{(n)\rho\sigma}
+\frac13 h_{\mu\nu}h^{\nu\rho}\Box h_\rho{}^\mu
+\frac14  hh_{\mu\nu}\Box h^{\mu\nu}).
\end{eqnarray}
As a convention, we set $\kappa$ as $\sqrt{16\pi G}$. 
The energy momentum tensor now becomes
$T_{\mu\nu}=\sum_n(\partial_\mu h_{\rho\sigma}^{(n)}\partial_\nu h^{(n)\rho\sigma})$. 
For the quantization of gravitational wave, 
the Hamiltonian is given by $H=\int d^3x~T^{00}$ then
\begin{eqnarray}
H
=\int\frac{d^3p}{(2\pi)^3}
\sum_{\lambda,n}
E_{\bm p}^{(n)}
a_{\bm p}^{(\lambda,n)\dagger}
a_{\bm p}^{(\lambda,n)},
\end{eqnarray}
where we omit the zero point energy. 
This is a familiar form and commutation relations are as usual except minus sings: 
$[H, a^{(\lambda,n)\dagger}_{\bm p}]
=(-1)^nE^{(n)}_{\bm p} a^{(\lambda,n)\dagger}_{\bm p}$. 
and $[\bm P, a^{(\lambda,n)\dagger}_{\bm p}]
=(-1)^n\bm p a^{(\lambda,n)\dagger}_{\bm p}$, 
where $\bm P=\int d^3 x T^{0i}$. 
Writing $P^\mu=(H,\bm P)$, they hold the identities 
$e^{-i P\cdot x}a_{\bm p}^{(\lambda,n)\dagger}e^{i P\cdot x}
=a_{\bm p}^{(\lambda,n)\dagger}e^{-(-1)^ni p\cdot x}$ 
and $e^{-i P\cdot x}a_{\bm p}^{(\lambda,n)\dagger}e^{i P\cdot x}
=a_{\bm p}^{(\lambda,n)\dagger}e^{-(-1)^ni p\cdot x}$. 
Thus, they provide the consistent specetime dependence of 
$h_{\mu\nu}^{(n)}(x)$.  
In addition, energies of ghost particles are positive \cite{Takook}.

Here, let us note the aspect of this quantization scheme. 
At first, the interaction part of the Hamiltonian violates 
Lorentz and CPT symmetries. In particular, 
matter-gravity couplings are important to see the effect of 
the violations. Various scenarios are sought to observe the 
Lorentz violation \cite{Kostelecky}. 
Other perturbative methods of quantum gravity 
such as the Lagrange formalism \cite{Dewitt} and the ADM action \cite{Bomstad} 
rarely violate the Lorentz symmetry so that it is a characteristic feature. 
Secondly, interaction terms appear for the momentum operator 
since $G^{0i}$ and $T^{0i}$ of matter sector 
in a curved spacetime are not zero in general. 
Adding to the time evolution, 
the configuration of the wave packet gains space evolution by the interaction 
via gravitons. 
It is not discussed in the literature and we do not know how it affects to particle physics. 
This effect may be observable as the size of wave packet is relating to the probability density. 
At any rate, the gravitational interaction is very weak so that 
it will not leave problematic result. 
Other important thing to discuss is how 
to derive mass, ghost kinetic, and cancellation terms. 
There needs more fundamental theory to produce them. 
Considering the theory including higher curvature terms, 
they will appear naturally and the quantization 
can be performed without adding terms by hand. 
However, higher curvature terms make stronger divergences 
and we cannot avoid fine-tuning to cancel out them. 
To derive the system discussing here, 
one will need more powerful theory.

\subsection{Propagator}
To calculate vacuum energy and graviton mass correction, 
let us consider concrete forms of propagators for small ${\cal N}$. 
In curved spacetime, vacuum is not static or not empty in general. 
We can define that the vacuum is empty at $t=0$ by 
$a_{\bm p}^{(\lambda,n)}|\Omega(0)\rangle=0$ but 
in a later time the background metric mixes the 
positive and negative frequency components \cite{Lovas}. 
However, calculations of quantum effects with empty vacuum writing $|0\rangle$ 
will be sufficient because we only try order estimation for latter sections. 

The propagator of empty vacuum can be calculated by
\begin{eqnarray}
\langle0|T( h_{\mu\nu}^{(n)}(x) h_{\rho\sigma}^{(n)}(y))|0\rangle
=(-1)^{n}\int\frac{d^4p}{(2\pi)^4}
\frac{-iP_{\mu\nu\rho\sigma}e^{(-1)^nip\cdot(x-y)}}{p^2+m^2_{n}-(-1)^ni\epsilon},
\end{eqnarray}
where $P_{\mu\nu\rho\sigma}
=\frac12(\eta_{\mu\sigma}\eta_{\nu\rho}
+\eta_{\mu\rho}\eta_{\nu\sigma}
-\frac23\eta_{\mu\nu}\eta_{\sigma\rho})$. 
From now on, we do not explicitly write $\epsilon$. 
As an example, 
the Feynman rule of the propagator for ${\cal N}=2$ is
\begin{eqnarray}
D_{\mu\nu\rho\sigma}=
\frac{-i(m_1^2-m_0^2) }
{(p^2+m^2_{0})(p^2+m^2_{1})}P_{\mu\nu\rho\sigma}.
\end{eqnarray}
This can make the theory renormalizable 
and the number of courter terms will be finite.

Since the purpose of this paper is to avoid fine-tuning, 
${\cal N}=2$ is not enough. 
To make a super-renormalizable model, we use $N=4$ 
with assuming $m_3=\sqrt{m_0^2-m_1^2+m_2^2}$, then 
\begin{eqnarray}
D_{\mu\nu\rho\sigma}=
\frac{-i(m_1^2-m_0^2)(m_1^2-m_2^2)
(m_0^2+m_2^2+2p^2)}
{(p^2+m^2_{0})(p^2+m^2_{1})(p^2+m^2_{2})(p^2+m^2_{3})}P_{\mu\nu\rho\sigma}.
\end{eqnarray}
In the same way, we can get the super-renormalizable model for 
any even ${\cal N}$ with the condition 
$m_{{\cal N}-1}=\sqrt{\sum_{n=0}^{{\cal N}-2}(-1)^nm_n^2}$. 
Giving one more condition for ${\cal N}\geq6$, the theory becomes finite. 
For instance, when ${\cal N}=6$, conditions for the finite-field theory are 
$m_0^2-m_1^2+m_2^2-m_3^2+m_4^2-m_5^2=0$ and 
$m_0^2(m_2^2+m_4^2)+m_2^2m_4^2-m_1^2(m_3^2+m_5^2)-m_3^2m_5^2$. 
In this case, the propagator is
\begin{eqnarray}
D_{\mu\nu\rho\sigma}
=\frac{-i(f_1+f_2p^2+f_3p^4)}
{(p^2+m^2_{0})(p^2+m^2_{1})(p^2+m^2_{2})(p^2+m^2_{3})(p^2+m^2_{4})(p^2+m^2_{5})}
P_{\mu\nu\rho\sigma},
\end{eqnarray}
where $f_1$, $f_2$, $f_3$ denote functions of masses. 
In the approximation $m_1\approx m_3$ and 
$m_0\approx 0$, it can be written as
\begin{eqnarray}
\frac{-i
(m_2^2-m_1^2)^2(p^2+m_1^2)(3(m_2^2-2m_1^2)p^2+2m_2^4-3m_1^2m_2^2)}
{(m_2^2-2m_1^2)^2(p^2+m^2_{0})(p^2+m^2_{1})(p^2+m^2_{2})(p^2+m^2_{3})(p^2+m^2_{4})(p^2+m^2_{5})}
P_{\mu\nu\rho\sigma},
\end{eqnarray}
where we used $m_2>m_1$. 
With this propagator, all loop calculations become finite 
and we do not need renormalization.

\section{Fermion interaction}

The extension of fermion field to be valid in curved spacetime 
is formulated via vielbein and covariant derivative. 
Vielbein can transform the generic metric tensor 
to the Minkowski metric, i.e. $g_{\mu\nu}=e_{\mu}{}^ae_{\nu}{}^b\eta_{ab}$. 
The connection for the orthonormal frame is written by 
$\omega_\mu^{ab}$. Using this connection, the covariant derivative 
which acts on a spinor is given by $D_\mu\psi=\partial_\mu\psi
+\frac{i}{4}\omega_\mu^{ab}\sigma_{ab}\psi$ where 
$\sigma^{ab}=-\frac12[\gamma^a,\gamma^b]$. 
We choose the notation of gamma matrices to be 
$\{\gamma^a,\gamma^b\}=-2\eta^{ab}$. 
As a general framework, we assume torsion is not vanished. 
Extending the theory based on the Cartan geometry, 
torsion and curvature can be implemented naturally. 

The Lagrangian of fermion field in curved spacetime becomes  
\begin{eqnarray}
{\cal L}_D
=\frac i2
(-(\overline{D_\mu \psi})\gamma^a e_{a}{}^{\mu }\psi
+\bar\psi\gamma^a e_{a}{}^{\mu }D_\mu\psi)
-m\bar\psi \psi.
\end{eqnarray}
The connection can be explicitly written as
\begin{eqnarray}
\omega_{\mu ab}=\frac12e_{a}{}^\nu(\partial_\mu e_{b\nu}-\partial_\nu e_{b\mu})
-\frac12e_{b}{}^\nu(\partial_\mu e_{a\nu}-\partial_\nu e_{a\mu})
+\frac12e_{a}{}^\rho e_b{}^\sigma(\partial_\sigma e_{c\rho}-\partial_\rho e_{c\sigma})e_\mu{}^c.
\end{eqnarray}
Since the energy momentum tensor is obtained by 
$T_{\mu\nu}=e_{a \nu}T^{a}{}_\mu
=-\frac{e_{a \nu}}{\sqrt{-g}}\frac{\partial ({\cal L}_D\sqrt{-g})}{\partial e_a{}^\mu}$ \cite{Kawai}, 
we have
\begin{eqnarray}
T^{\mu\nu}
=\frac i2(\overline{D^\nu \psi })\gamma^a e_a{}^\mu 
\psi
-\frac i2\bar\psi\gamma^a e_a{}^\mu D^\nu \psi.
\end{eqnarray}
This is asymmetric if there is a torsion. 
Around the Minkowski space background, vielbein can be expanded as 
$e_a{}^\mu=\delta_a{}^\mu-\frac\kappa2h_a{}^\mu+\cdots$ 
and $e_{a\mu}=\eta_{a\mu}+\frac\kappa2h_{a\mu}+\cdots$. 
Then the first order of the Hamiltonian density is
\begin{eqnarray}
{\cal H}_\text{int}
=\frac {i\kappa}4(\partial^0\bar \psi)\gamma^a h_{a}{}^0\psi
-\frac {i\kappa}4\bar \psi\gamma^a h_{a}{}^0\partial^0\psi
+\frac\kappa{16}(\partial_a h^0{}_{b}-\partial_b h^0{}_{a})
\bar\psi(\gamma^0\sigma^{ab}+\sigma^{ab\dagger}\gamma^0)\psi.
\end{eqnarray}
It can make the Feynman rule of
\begin{eqnarray}
 \parbox{25mm}{
\includegraphics[width=2.9cm,height=1.9cm]{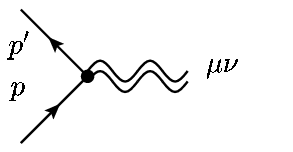}
}\hspace{1mm}
=-\frac {i\kappa}8(E+E')
(\gamma^\mu\eta^{\nu0}+\gamma^\nu\eta^{\mu0})
-\frac{\kappa}{4i}\epsilon_{i jk}
(\eta^{\mu i}\eta^{\nu0}+\eta^{\nu i}\eta^{\mu0})
(p^{j}-p'^{j})\Sigma^k,
\end{eqnarray}
where $\Sigma_k=\sigma_k\otimes \bf 1_{2\times 2}$.

\subsection{Newtonian gravity}

In the non-relativistic limit, $(0,0)$ component of 
the vertex of fermion-fermion-graviton is dominant  then 
\begin{eqnarray}
\label{Newtonian}
 \parbox{70mm}{
\includegraphics[width=2.9cm,height=1.9cm]{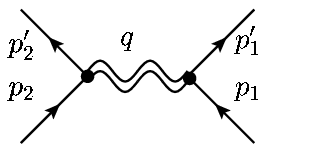}
}\hspace{-37mm}
\approx
\frac{i\kappa^2M_1M_2 P_{0000}u^\dagger(p_2') u(p_2)u^\dagger(p_1') u(p_1)}{4}
\sum_{n=0}^{{\cal N}-1}
\frac{ (-1)^n}{q^2+m_0^2},
\end{eqnarray}
where $M_1$ and $M_2$ are masses of fermions. 
The calculation of the gravitational potential is straightforward:
\begin{eqnarray}
\label{potential}
V(r)
=-\frac{\kappa^2M_1M_2}{12\pi r}
\sum_{n=0}^{\cal N}(-1)^ne^{-m_nr}.
\end{eqnarray}
Since $m_0$ is quite light while others are much heavier, 
it can be approximated by $V(r)\approx-\frac{\kappa^2M_1M_2}{12\pi r}$ 
which is $4/3$ times larger 
than Newtonian gravity when $G$ is the observed gravitational constant. 
This is the famous problem known as the vDVZ discontinuity.

To see the problem in our framework, let us consider the Einstein equation (\ref{eom}). 
It is sufficient to deal only with $h^{(0)}_{\mu\nu}$ since 
other gravitons do not influence in macroscopic scale. 
The equation of motion of lightest graviton is
\begin{eqnarray}
\begin{split}
(\Box-m_0^2) h^{(0)}_{\mu\nu}
=-\frac{16\pi G}{\kappa} 
(T_{\mu\nu}-\frac13T\eta_{\mu\nu}),
\end{split}
\end{eqnarray}
where we neglect gauge dependent term 
$\partial_\mu\partial_\nu/m_0^2$. 
Considering a point source with 
$T_{\mu\nu}=M\eta_{\mu0}\eta_{\nu0} \delta^3(\bm x)$, 
general solution is 
\begin{eqnarray}
h_{\mu\nu}^{(0)}
=\frac{16\pi G}{\kappa}
\int\frac{d^4p}{(2\pi)^4}
\frac{2\pi Me^{i\bm p\cdot x}\delta(p^0)}{\bm p^2+m^2_0}
(\eta_{\mu0}\eta_{\nu0}
+\frac{1}{3}\eta_{\mu\nu}).
\end{eqnarray}
Then
\begin{eqnarray}
h_{00}^{(0)}
=\frac{16\pi G}{\kappa}
\frac{2 Me^{-m_0 r}}{12\pi r},
\quad
h_{ij}^{(0)}
=\frac{16\pi G}{\kappa}
\frac{Me^{-m_0 r}}{12\pi r}\delta_{ij}.
\end{eqnarray}
For the light bending, the angle of deflection can be estimated as
\begin{eqnarray}
\alpha
\approx\frac{4 GM}{ R},
\end{eqnarray}
where $R$ is the closest approach to the source. 
This is the same result of general relativity, 
then we cannot change $G$ to fit the gravitational potential (\ref{potential}).

Commonly, it is considered that nonlinear effect of quantum correction can recover the 
smooth connection of massless and massive gravitational theories. 
Leading contribution (\ref{Newtonian}) is reliable only when 
the distance is longer than the Vainshtein radius which is defined by 
$(m^{-4}_0R_S)^{1/5}$ where $R_S$ is the Schwarzschild radius \cite{Babichev}. 
This becomes quite long if the mass is extremely small, 
e.g., considering $m_0$ is the inverse of the Hubble constant, 
the Vainshtein radius of the Sun becomes about $100$kpc 
which is longer than the size of the Milky way. 
Then nonlinearity is important for the region of the solar system. 
Summing up all the quantum corrections, the result may be correspondent to 
the case of massless graviton (see e.g. \cite{Deffayet}). 
However, this scenario can be applied when corrections from 
loop diagrams are not small compared to the leading order. 
If usual propagator is used and cutoff scale is the Planck mass, 
loop corrections are comparable. 
In our case, ghost gravitons suppress the corrections so that 
nonlinear effect is not important. 
That means the problem is serious and the theory is inconsistent with experiments  
unless we can find other remedy.

As noted in previous section, 
there may be more fundamental theory that may lead all additional terms of the Lagrangian. 
If the fundamental theory does not involve the problem of 
the vDVZ discontinuity, the theory will be correspondent to the model of massless graviton. 
For instance, basing higher-derivative gravity without Pauli-Fierz mass, 
the discontinuity does not exist \cite{Nakasone}. 
By modifying kinetic terms or strengths of interaction terms, 
we can avoid the problem. 
A simple example is to take
\begin{eqnarray}
\begin{split}
{\cal L}_\text{vDVZ}
=&-\frac{1}{8\sqrt{-g}}
(h^{(0)\mu\nu}\Box h_{\mu\nu}^{(0)}
-2h^{(0)}_{\mu\nu}\partial^\mu\partial_\sigma h^{(0)\sigma\nu}
+2h^{(0)}_{\mu\nu}\partial^\mu\partial^\nu h^{(0)}
-h^{(0)}\Box h^{(0)}))
\\
&+\frac{1}{8\sqrt{-g}}
m_0^2(h^{(0)}_{\mu\nu}h^{(0)\mu\nu}-h^{(0)}h^{(0)}).
\end{split}
\end{eqnarray}
They can change the equations of motion so that the angle becomes 
$\alpha\approx\frac{16GM}{ 3R}$. 
With this modification, the gravitational force and light bending 
are consistent with experiments 
when $G=\frac34G_N$ where $G_N$ is the measured gravitational constant.

\section{Ghost and unitarity}

Ghost particles generally violate unitarity because of negative norms. 
There are three conditions to preserve unitarity \cite{Kugo}: 
(i) the $S$-matrix of whole state space ${\cal V}$ is unitary, 
(ii) physical space ${\cal V}_\text{phys}$ which is a subset of 
${\cal V}$ is invariant under the $S$-matrix, namely 
$S{\cal V}_\text{phys}=S^\dagger{\cal V}_\text{phys}={\cal V}_\text{phys}$, 
(iii) physical space has positive semi-definite metric, 
for any $|\text{phys}\rangle \in {\cal V}_\text{phys}$,  
we should hold $\langle\text{phys}|\text{phys}\rangle\geq 0$. 
Even when ghost particles are added, all of them can be satisfied by assuming the mixing 
state of normal and ghost particles as a fundamental set. 
Additionally, specific wave packet is assumed to satisfy the unitarity.

\subsection{With renormalizable condition}
We first consider ${\cal N}=2$ so the metric is 
$h_{\mu\nu}=h_{\mu\nu}^{(0)}+h_{\mu\nu}^{(1)}$. 
In the analogy of longitudinal polarization of photon, 
we will try to cancel negative norms by constraining the system. 
Suppose the Gupta-Bleuler quantization 
which gives the constraint $(a_{\bm p}^{(0)}-a_{\bm p}^{(3)})|\text{phys}\rangle=0$ 
where $a_{\bm p}^{(n)}$ are annihilation operators of photon, 
the same constraint for the creation operators of gravitons will be useful 
$(a_{\bm p}^{(0,\lambda)}+a_{\bm p}^{(1,\lambda)})|\text{phys}\rangle=0$. 
In this constraint, the ghost graviton is always appeared with 
the normal graviton, then the negative norm will not appear. 
This is the key to satisfy the third condition of unitarity. 
For the case of gauge group, general proof of unitarity is given by 
\cite{Kugo} and it will be partly applicable for the case here.

The constraint for the physical state will differ from the 
one of photon due to different masses and wave functions. 
To be more precise, we define
\begin{eqnarray}
|h_{\mu\nu}^{(n)}\rangle
=\frac{1}{\sqrt5}\int \frac{d^3p}{(2\pi)^3}
\sum_\lambda 
e^{(\lambda)}_{\mu\nu}
a^{(\lambda,n)\dagger}_{\bm p}\Phi_{\bm p}^{(n)}
|0\rangle,
\end{eqnarray}
where wave packets $\Phi^{(n)}_{\bm p}$ are assumed to be 
$\int d^3p~\Phi_{\bm p}^{(n)\dagger}\Phi_{\bm p}^{(n)}=1$ so that 
$\langle h^{(n)}_{\mu\nu}|h^{(m)\mu\nu}\rangle=(-1)^{n+1}\delta^{nm}$. 
Using them, we get the zero-norm state 
$|h_{\mu\nu}\rangle=\frac1{\sqrt2}
(|h_{\mu\nu}^{(0)}\rangle+|h_{\mu\nu}^{(1)}\rangle)$ where 
the normalization factor is put to be compared with positive-definite norms. 
The state $|h_{\mu\nu}\rangle$ can be a physical state if 
wave packets $\Phi^{(0)}_{\bm p}$ and $\Phi^{(1)}_{\bm p}$ have the relation 
$\int d^3p(\Phi^{(0)}_{\bm p}-\Phi^{(1)}_{\bm p})=0$. 
Then the wave packets of normal and ghost particles will be the same. 
The constraint is rather simple and seems reasonable if the cancellation 
works between them, nonetheless, the problem is the dependence on the $S$-matirx. 
As it cannot satisfy the conditions of unitarity, 
the constraint needs to be modified. Alternatively, we take 
$(h^{(0,+)}_{\mu\nu}+h^{(1,+)}_{\mu\nu})|\text{phys}\rangle=0$ where 
\begin{eqnarray}
h^{(n,+)}_{\mu\nu}
=\int\frac{d^3p}{(2\pi)^3}
\frac{1}{\sqrt{2E_{\bm p}^{(n)}}}
\sum_\lambda e_{\mu\nu}^{(\lambda)}
a_{\bm p}^{(\lambda,n)}.
\end{eqnarray}
The new constraint can realize the independence on the $S$-matrix 
since any interaction appears with $h_{\mu\nu}^{(0)}+h_{\mu\nu}^{(1)}$. 
If $|h_{\mu\nu}\rangle$ is a physical state, wave packets are related with
$\int d^3p(\Phi^{0}_{\bm p}/\sqrt{E_{\bm p}^{(0)}}
-\Phi^{1}_{\bm p}/\sqrt{E_{\bm p}^{(1)}})=0$. 
When the wave packets are Gaussian 
$\Phi_{\bm p}^{(n)}=\frac{e^{-\bm p^2/2\omega_n^2}}{(\sqrt{\pi}\omega_n)^{3/2}}$, 
the relation becomes
\begin{eqnarray}
\label{const}
{\omega_0U[\frac14,-\frac14,\frac{m_0^2}{2\omega_0^2}]
-\omega_1U[\frac14,-\frac14,\frac{m_1^2}{2\omega_1^2}]}=0,
\end{eqnarray}
where $U[a,b,z]$ is the confluent hypergeometric function 
of the second kind. Assuming $m_0/\omega_0\ll1$ and 
$m_1/\omega_1\ll1$, we get 
\begin{eqnarray}
\omega_1\approx
\frac{4\sqrt2\Gamma^2[\frac54]\omega_0^2}{\pi m_1}.
\end{eqnarray}
In this way, we can get $S|h_{\mu\nu}\rangle=0$ by the constraint of wave packets.

Usually, time-independence is supposed as a necessary condition 
for the conservation of unitarity. In fact, if there is a time evolution 
of $|h_{\mu\nu}\rangle$, there appears new state 
$|h'_{\mu\nu}\rangle=|h_{\mu\nu}^{(0)}\rangle-|h_{\mu\nu}^{(1)}\rangle$ 
and the state of gravitational field turns out to be 
the mixing of $|h_{\mu\nu}\rangle$ and $|h'_{\mu\nu}\rangle$. 
Similar to the neutrino oscillation, this phenomenon occurs 
since normal and ghost gravitons have different masses. 
Although both states have zero norms, 
their product $\langle h_{\mu\nu}|h'^{\mu\nu}\rangle$ 
is negative. Then the appearance of whichever $|h_{\mu\nu}\rangle$
or $|h'_{\mu\nu}\rangle$ is problematic. 
Once one can exist, the other is derived by the oscillation. 
However, it does not a matter since the probability to produce 
$|h_{\mu\nu}\rangle$ is zero while $|h'_{\mu\nu}\rangle$ cannot satisfy the constraint 
of physical state. 
Therefore the problem can be avoided if we add the constraint that 
$|h_{\mu\nu}\rangle$ does not exist in the initial condition.

A proof of unitarity conservation will follow by subtracting zero-norm 
states from the entire physical space if they do not affect for physics. 
After this, only positive definite metrics remain. 
Nonetheless, a confusion may occur since the energy of the state $|h_{\mu\nu}\rangle$ 
is not zero. The energy is calculated by
\begin{eqnarray}
\label{ene2}
\langle h_{\mu\nu}|H|h^{\mu\nu}\rangle
=\frac12
\left(
\frac{e^{m^2_0/2\omega_0^2}m_0^2
K_1[\frac{m_0^2}{2\omega_0^2}]}{\sqrt\pi\omega_0}
+\frac{e^{m^2_1/2\omega_1^2}m_1^2
K_1[\frac{m_1^2}{2\omega_1^2}]}{\sqrt\pi\omega_1}\right),
\end{eqnarray}
where $K$ is the modified Bessel function of the second kind. 
This energy will be related to when the constraint of physical state begins. 
Below the energy scale, there is no problem 
since $|h_{\mu\nu}^{(0)}\rangle$ has positive 
norm and the ghost $|h_{\mu\nu}^{(1)}\rangle$ does not appear. 
In addition, $|h_{\mu\nu}^{(0)}\rangle$ indicates the ordinary gravitational wave 
which is indirectly observed. 
Therefore, it is necessary to assume that the physical state of the graviton 
is first $|h_{\mu\nu}^{(0)}\rangle$ and later replaced by 
$|h_{\mu\nu}\rangle$ when the energy is beyond the scale 
that $|h_{\mu\nu}\rangle$ can appear in the real world. 
The exact value of the lowest energy should be calculated by 
varying $\omega_0$ and $\omega_1$ with the constraint of Eq. (\ref{const}). 
A subtlety remains since the energy changes with the normalization 
factor which cannot be determined due to zero norm. 
In other words, the zero-norm state affects for the observation indirectly 
that is completely different from the case of longitudinal photon. 
For now, as the determination of the factor seems not a crucial issue, 
we just postulate that normalizations of zero norms are 
the same to those of positive definite metrics and then 
there is no ambiguity. 

The phenomenon of the change of the state from the one of gravitational wave 
to the zero-norm state makes important prediction. 
Since $S$-matrix with zero-norm states become trivial, 
all the probabilities including the states are zero. 
Therefore, the process that increases the energy of gravitational 
wave beyond the energy about $m_1/2$ when the state is changed to 
$|h_{\mu\nu}\rangle$ will be canceled out. 
That implies the energy of gravitational wave will have the upper bound. 
Concretely, there is no way to increase the energy beyond the bound. 
This prediction is not so significant for gravitons, but if we apply the 
same scenario for other elementary particles like electron and neutrino, 
it is noteworthy. As discussed in section 6, 
particles of the standard model may have ghost partners so that 
the prediction to have upper energy bound will be interesting.

\subsection{With super-renomalizable condition}

We use the same techniques for ${\cal N}>2$ with conditions which can 
lead higher power of propagator.  
For instance, when ${\cal N}=4$, 
the super-renormalization condition is $m_3=\sqrt{m_0^2-m_1^2+m_2^2}$. 
In this case, the mass $m_2$ should be heavier than $m_3$ as $m_0<m_1$. 
Then, we cannot use the cancellation of the negative norm of  
$|h_{\mu\nu}^{(2)}\rangle$ by constraining to 
the zero norm $\frac1{\sqrt2}(|h_{\mu\nu}^{(2)}\rangle+|h_{\mu\nu}^{(3)}\rangle)$. 
Instead, we require two steps for the physical state: 
the first is the same to ${\cal N}=2$, i.e. 
$(h_{\mu\nu}^{(0,+)}+h_{\mu\nu}^{(1,+)})|\text{phys}\rangle=0$ 
at the scale larger than the minimum of 
Eq. (\ref{ene2}) with the state 
$\frac{1}{\sqrt2}(|h_{\mu\nu}^{(0)}\rangle+|h_{\mu\nu}^{(1)}\rangle)$, 
the next is to replace this state by 
$|h_{\mu\nu}\rangle=\frac{1}{\sqrt4}
\sum_{n=0}^3|h_{\mu\nu}^{(n)}\rangle$ from its lowest energy 
about $(m_1+m_2+m_3)/4$ with the constraint 
$\sum_n h_{\mu\nu}^{(n,+)}|\text{phys}\rangle=0$. 
To suppress the ghost state $|h^{(3)}_{\mu\nu}\rangle$, 
a condition for a mass appears, precisely, 
$m_3$ should be larger than the lowest energy of $|h_{\mu\nu}\rangle$. 
It can be calculated by searching the minimum of
\begin{eqnarray}
\langle h_{\mu\nu}|H|h^{\mu\nu}\rangle
=\frac14\sum_{n=0}^{3}\frac{e^{m^2_n/2\omega_n^2}m_n^2
K_1[\frac{m_n^2}{2\omega_n^2}]}{\sqrt\pi\omega_0},
\end{eqnarray}
with varying the parameters $\omega_n$ 
and taking the condition 
$\sum_n(-1)^n\omega_nU[\frac{1}{4},-\frac{1}{4},\frac{m_n^2}{2\omega_n^2}]=0$. 
When $m_2$ and $m_3$ are similar values, $\omega_2$ 
and $\omega_3$ are approximately equal. 
Then the constraints for the masses can be approximated as
$0<m_1<m_2/\sqrt2$ and $m_2<m_3$.

Let us discuss in more general context. 
The super-renormalization condition for general ${\cal N}$ is 
$m_{{\cal N}-1}=\sqrt{\sum (-1)^nm_n^2}$. 
Ghost masses are assumed to be larger as the number is increasing, i.e.
$m_{2n-1}<m_{2n+1}$ for any $n$. 
Defining $|h_{\mu\nu}^{[2N]}\rangle
=\frac{1}{\sqrt{2N}}\sum_{n=0}^{2N-1}|h_{\mu\nu}^{(n)}\rangle$, 
we requre that states are changing in a stepwise fasion from 
$|h_{\mu\nu}^{(0)}\rangle$ to $|h_{\mu\nu}^{[2]}\rangle$, 
$\cdots$, $|h_{\mu\nu}^{[{\cal N}]}\rangle=|h_{\mu\nu}\rangle$ 
depending on the energy scale where each state can appear. 
A constraint for the physical state $|h_{\mu\nu}^{[{\cal N}']}\rangle$ is given by 
$\sum_{n=0}^{{\cal N}'}h_{\mu\nu}^{(n,+)}|\text{phys}\rangle=0$. 
All ghosts masses should have 
$m_{2N+1}>\frac{1}{2N+2}\sum_{n=0}^{2N+1}m_{n}$ to cancel each negative norm by 
involved in zero-norm states. 
Except $|h_{\mu\nu}^{(0)}\rangle$, norms are zero and 
unaffected by the $S$-matrix so that nothing appears in 
the physical world. 

For individual ghost state, there is at least one condition 
to be involved in zero-norm states, on the other hand, 
normal gravitons do not need to be a part of such states. 
For instance, the mass of $h_{\mu\nu}^{(4)}$ in ${\cal N}=6$ 
can be lighter than the lowest energy of $|h_{\mu\nu}\rangle$. There is no other constraint 
for $h_{\mu\nu}^{(4)}$ then $|h_{\mu\nu}^{(4)}\rangle$ 
can appear in the final state. The behavior of this particle is uncommon because the 
available momentum is limited, namely, the range is from $m_4$ to $\frac16\sum m_n$. 
Depending on a parameter region, 
the motion of $h_{\mu\nu}^{(4)}$ can be restricted only in non-relativistic regime. 
Further, as it is neutral from any gauge interaction, life-time will be long enough. 
Then it is a new candidate of a dark matter.

When ${\cal N}\geq 6$, all quantum effects become finite by the two conditions: 
$m_{{\cal N}-1}=\sqrt{\sum_{n=0}^{{\cal N}-2}(-1)^nm_n^2}$ 
and $\sum_{m=0}^{{\cal N}-1}(-1)^m(-\sum_{n=0}^{{\cal N}-1}m_m^2m_n^2+m_m^4)=0$. 
Even in this case, the above scheme can be applied to preserve 
unitarity with the constraints $\sum_n h_{\mu\nu}^{(n,+)}|\text{phys}\rangle=0$ 
and $m_{2N+1}>\frac{1}{2N+2}\sum_{n=0}^{2N+1}m_{n}$. 
The allowed region of mass parameters will be more constrained, 
but it is possible to obtain the physical state. 
This will be useful to make quantum effects significantly suppressed.

\section{Vacuum energy}
The vacuum energy diverges in the standard calculation of quantum field theory and 
it induces the cosmological constant problem. 
Using the ghost particle, the propagator gains 
higher power so that the divergence becomes weaker. 
Using the cutoff scale $\Lambda$, the vacuum energy without ghost is of order $\Lambda^4$ 
and it can be improved by ghost particles with the super-renormalization condition 
into the value of order the fourth power of the ghost masses. 
Experimentally, the cosmological constant is observed about $29\text{meV}^4$. 
As we can insert any value for the ghost mass, 
it is not difficult to lead the observed value. 
However, additional gravitons enforce to 
change the gravitational potential to 
$\frac{G_NM_1M_2}{r}(1+\sum(-1)^ne^{-nm_n r})$. 
The strongest bound of the Newtonian gravity for the type 
$\frac{G_NM_1M_2}{r}(1+e^{-m_y r})$ is roughly 
$m_y>1/20\mu\text{m}$ \cite{Rajalakshmi}. 
Then, we cannot use the mass scale derived from the cosmological constant 
$\lambda^{1/4}\approx1/85\mu $m 
despite the natural choice to be consistent with the observation. 
Actually, it is not an easy task to find allowed parameter region satisfying 
all the constraints of experiments and unitarity. 
We abandon the complete analysis but show an example of 
consistent solution.

The vacuum energy can be calculated by
\begin{eqnarray}
T^{\mu\nu}_\text{vac}=\frac{1}{8\pi G}\langle0|G^{\mu\nu}|0\rangle.
\end{eqnarray}
To the leading order, it becomes
\begin{eqnarray}
T^{\mu\nu}_\text{vac}
=\lim_{x-y\rightarrow0}\sum_n
\langle0|T(\partial^\mu h_{\rho\sigma}^{(n)}(x)\partial^\nu h^{(n)\rho\sigma}(y))|0\rangle.
\end{eqnarray}
This can be rewritten as
\begin{eqnarray}
T^{\mu\nu}_\text{vac}
=\lim_{x-y\rightarrow0}
\sum_n\int \frac{d^4p}{(2\pi)^4}
\frac{-i(-1)^np^\mu p^\nu}{p^2+m_n^2-(-1)^ni\epsilon}e^{ip\cdot(x-y)}.
\end{eqnarray}
When ${\cal N}=2$ and the momentum cutoff is taken, 
it is integrated into 
\begin{eqnarray}
T^{00}_\text{vac}
=\frac{8(m_1^2-m_0^2)\Lambda^2}{128\pi^3},
\quad
T^{ii}_\text{vac}
=-\frac{8(m_1^2-m_0^2)\Lambda^2}{384\pi^3}.
\end{eqnarray}
Considering $\Lambda$ is the Planck scale, 
$m_1^2-m_0^2$ should be ${\cal O}(10^{-38})$GeV. 
Apparently, $m_1$ cannot be so small to realize 
large-scale gravity. 
On the other hand, if we consider ${\cal N}=4$ with the condition of super-renormalization, 
the energy-momentum tensor becomes
\begin{eqnarray}
T^{\mu\nu}_\text{vac}
=\int \frac{d^4p}{(2\pi)^4}
\frac{-i(m_1^2-m_0^2)(m_1^2-m_2^2)(m_0^2+m_2^2+2p^2)p^\mu p^\nu}
{(p^2+m_0^2)(p^2+m_1^2)(p^2+m_2^2)(p^2+m_3^2)}.
\end{eqnarray}
The leading contribution is proportional to the logarithm of $\Lambda$:
\begin{eqnarray}
T^{\mu\nu}_\text{vac}
=\frac{m_1^2(m_2^2-m_1^2)}{16\pi^3}\ln[\frac{\Lambda}{m_2}]\eta^{\mu\nu}
+\cdots.
\end{eqnarray}
It has a wrong sign and the vacuum energy becomes negative. 
This is because the contribution from ghost particles is larger than 
the one of normal gravitons. Although ${\cal N}=4$ is the least number to be 
super-renormalizable, it cannot yield a consistent result.

The simplest predictive model is ${\cal N}=6$ with the super-renormalization condition. 
To get a rough estimation, we take 
$m_0\approx0$ and $m_2\approx m_4$, 
then the energy-momentum tensor in vacuum becomes
\begin{eqnarray}
\label{ene}
T^{\mu\nu}_\text{vac}
\approx -i
\int\frac{d^4p}{(2\pi)^4}
\frac{-2((m_2^2-m_5^2)^2-m_1^2m_3^2)p^4
+3m_1^2m_3^2m_5^2(p^2+m_2^2)}
{(p^2+m_1^2)
(p^2+m_2^2)(p^2+m_3^2)(p^2+m_5^2)}\eta^{\mu\nu}.
\end{eqnarray}
The leading is
\begin{eqnarray}
\begin{split}
\label{func}
T^{\mu\nu}_\text{vac}
\approx
\frac{-4\pi^2}{(2\pi)^4}
((m_1^2+m_3^2-m_2^2)^2-m_1^2m_3^2)\ln[\frac{\Lambda}{m_2}]\eta^{\mu\nu}
+\cdots.
\end{split}
\end{eqnarray}
This time, it is possible to keep a positive energy within the constraint of unitarity. 
In order to get $T^{00}\approx 29\text{meV}^4$, 
masses are estimated around $3$meV if they have the same order without parameter tuning. 
On the other hand, if we tune the parameters e.g. $m_1\approx m_3\approx m_2/\sqrt3$, 
masses can be in any order depending on the level of tuning. 
More concretely, giving masses around the condition of finite-field theory, 
the magnitude of the leading contribution can be suppressed. 
For the general case (\ref{ene}), we performed numerical analysis 
to find that all masses can exceed $20$meV and Fig. 1 (left) exhibits some parameter sets. 
Taking a set of masses, we predict the modified Newtonian gravity by
\begin{eqnarray}
V(r)
\approx
-\frac{G_NM_1M_2}{r}
(e^{m_0r}-e^{m_1r}+e^{m_2r}-e^{m_3r}+e^{m_4r}-e^{m_5r}).
\end{eqnarray}
One interesting example is $(m_1,m_2,m_3,m_4,m_5)\approx
(21.3,37.1,21.4,30.5,37.3)\text{meV}$. 
Compared with the Yukawa-type modification, 
the effective mass is approximately $1/14$-$1/10~\mu\text{m}^{-1}$ 
which is shown in Fig. 1 (right). 
Increasing the tuning level, larger mass sets will appear whereas 
they are less interesting since the problem of lightest graviton mass 
arises as discussed in the next section.

The cosmological constant problem is much improved compared to the fine-tuning solution 
without ghost particles. 
However, if future experiments verify the inverse square law in shorter range, 
we need to increase the tuning level depending on the range. 
Strictly speaking, as the scale $85\mu$m of natural prediction 
is already rejected, 
it does not solve the cosmological constant problem at all. 
Using larger ${\cal N}$, we can take the finite-field conditions exactly, 
but similar tuning will be necessary. 
It may be important to consider additional mechanism 
or symmetry to reduce the amount of vacuum energy.

\begin{figure}[t]
\includegraphics[width=7cm,height=4.2cm]{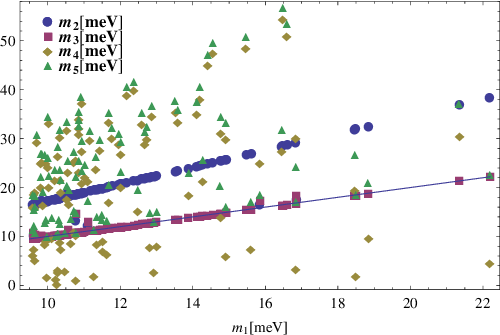}
\qquad
\includegraphics[width=7cm,height=4.2cm]{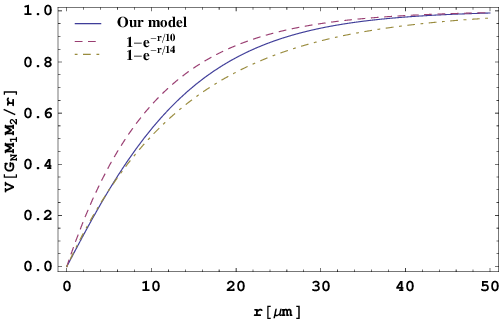}
\caption{Left figure shows the parameters sets 
which satisfy the cosmological constant and the constraints of unitarity. 
The solid line indicates the mass of $m_1$. 
Large masses can appear when 
the conditions of finite-field theory are approximately satisfied 
since they suppress the leading order proportional to the logarithm of the cutoff scale 
which is taken as the Planck mass. 
In right figure, we describe the modified gravity at short distance 
using one parameter set with relatively large masses obtained from the left figure. 
It is compared with the Yukawa-type potential.}
\end{figure}

\section{Graviton mass}

There are many models that predict modification of gravity in 
short range by using extra dimension. In our model 
with ghost particles, it can be distinguished from them 
by predicting modified Newtonian gravity at the cosmological scale via lightest graviton mass. 
From the loop calculations, parameters of gravitons masses are related to the lightest mass. 
The estimation of lightest mass is also important as it is near to the experimental bound. 
Conversely, if the mass of lightest graviton can be measured 
by observation, the parameter set of other gravitons masses can be restricted. 
After finding modified gravity in either short or long range, 
the model would make stronger prediction.

The most important effect of loop calculation of graviton mass correction 
comes from the four-gravitons vertex. 
The Feynman rule of the vertex is given by 
\begin{eqnarray}
{\cal H}_\text{int}^{(4)}
=-\frac12\kappa^4
\partial_\mu (h^{\mu\nu}h_{\nu\rho}
h^{\rho\sigma})\partial_0 h_{\sigma0}+\cdots.
\end{eqnarray}
The Feynman rule of the first term is estimated as
\begin{eqnarray}
 \parbox{60mm}
{\includegraphics[width=2.9cm,height=1.9cm]{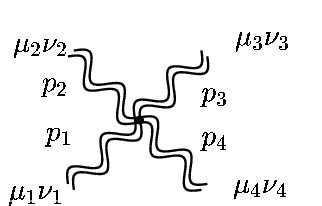}}
\hspace{-32mm}
=\frac1{4!}(-\frac1{2}\kappa^2
(p_2+p_3)^{\mu_1}
 (p_4)^0
\eta^{\nu_1\mu_2}
\eta^{\nu_2\mu_3}
\eta^{\nu_3\mu_4}
\eta^{\nu_40}
+1\leftrightarrow2\leftrightarrow3\leftrightarrow4).
\end{eqnarray}
Using this rule, the graviton mass correction becomes
\begin{eqnarray}
\Delta m^2
\sim-\frac{i}2\kappa^2\int\frac{d^4p}{(2\pi)^4}
p_0^2\sum_{n=0}^{\cal N}\frac{(-1)^n}{p^2+m_n^2}.
\end{eqnarray}
The estimation of whole diagrams is difficult and we do not try here. 
In quantum gravity without ghosts, 
it becomes $\Delta m^2\sim\Lambda^2$ which needs strong fine-tuning. 
When ${\cal N}=2$, it is $\Delta m^2\sim \kappa^2\Lambda^2 m_1^2$. 
This case also needs fine-tuning to make the graviton mass light to 
be consistent with the solar system. 
To maintain the gravity for the scale larger than 1pc, 
$\Delta m$ should be less than $10^{-20}$meV.

If a model is super-renormalizable, 
we can get a realistic graviton mass. 
The correction for ${\cal N}=4$ is
\begin{eqnarray}
\Delta m^2
\sim-\frac12\kappa^2\int\frac{d^4p}{(2\pi)^4}
\frac{-2im_1^2(m_1^2-m_2^2)p^2p_0^2}
{(p^2+m_0^2)(p^2+m_1^2)(p^2+m_2^2)(p^2+m_3^2)}.
\end{eqnarray}
Then the propagator becomes
\begin{eqnarray}
D_{\mu\nu\rho\sigma}
\sim\frac{-iP_{\mu\nu\rho\sigma}m_1^2(m_1^2-m_2^2)(m_2^2+p^2)}
{p^2(p^2+m_1^2)(p^2+m_2^2)(p^2+m_3^2)
+m_1^2(m_1^2-m_2^2)(m_2^2+p^2)\Delta m^2}.
\end{eqnarray}
Since $\Delta m^2\sim \frac{1}{8\pi^2}\kappa^2 m_1^4\ln[\Lambda^2/m_1^2]$, 
it is easy to get $\Delta m<1\text{pc}^{-1}$ when ghost masses are chosen 
to fit the cosmological constant. 
Inserting $m_1=\lambda^{1/4}$ and $\Lambda$ the Planck mass, 
it is of order $4\times 10^{-31}$meV 
or $6\times10^{-11}\text{pc}^{-1}$. 
The calculation for the case ${\cal N}=6$ is also straightforward. 
In previous section, one possible solution is given as example. 
Using the parameter set, we found $\Delta m\sim 1.4\times10^{-28}$meV 
or $2.2\times 10^{-8}\text{pc}^{-1}$. 
Since the case of higher-derivative theory is independent from 
the DGP model (Dvali-Gabadadze-Porrati), 
the graviton mass limit is $\lambda_g \geq 100$Mpc \cite{Goldhaber}. 
Then the solution of the model is marginal to the experiments. 
Taking the parameter set which has larger graviton masses, 
it will be inconsistent with the observation of the cosmological structure. 
Hence, ghost masses are favored to be in the range of 
10-100meV to suppress the lightest graviton mass. 
Above calculations reflect just one contribution of the quantum correction and 
actually there are thousands of contributions from similar terms and diagrams. 
It is inferred that by summing up other contributions 
lightest mass exceeds the experimental bound, then 
we will need some tuning with the bare mass $m_0$. 
The model does not indicate strong predictions 
for modified scales of gravity but implies favored ranges of the modification.

Lastly, let us comment about divergences 
induced by interactions with the matter sector. They are usually quartic divergences 
and non-renormalizable. There need infinite counter terms 
to cancel all the divergences even when higher curvature terms are added. 
One possible solution is to consider supergravity \cite{Bern} and 
it is known that supergravity can reproduce finite results for up to three-loop diagrams. 
But in general loop, no one knows whether the result can be finite. 
A different approach is to consider higher-derivative propagator for all the matter fields 
then it can make the theory renormalizable with power counting arguments. 
Taking the same procedure of section 4, unitarity can be preserved 
when ghost particles are introduced for any other particle. 
A critical difference is that particles in the standard model are investigated 
accurately for high energy so that ghost masses must be very large. 
For example, cosmic ray experiments of neutrino 
find that neutrino can have the energy $10^{12}$GeV, then the mass of ghost partner of 
neutrino must be larger than this value. This mass scale is too high to render the 
loop correction of graviton mass small enough. 
Although the method is more powerful than supergravity, 
it is still impossible to derive the consistent mass correction of 
lightest graviton without the fine-tuning. 
There is also a problem when the vacuum energy is calculated 
for the particles of the standard model. If ghost masses substantially exceed 
the energy scale of milli-electron volt, the cosmological constant cannot be explained. 
Thus, there needs more powerful method that can suppress 
the vacuum energy and the graviton mass correction when matter sector is added. 
For instance, by combining higher-derivative theory and supergravity, 
some strong prediction may be found. 
In this vein, it there is a strong tool, one can predict modified gravity 
via the quantum effects.

\section{Conclusion}
We have invented new techniques that can preserve unitarity 
for the system including ghost partners. 
They are so powerful that quantum gravity with matter fields can be 
renormalizable if ghost partners are introduced for every quantum field. 
Prediction of modified gravity in both short and long ranges is present 
for the model ${\cal N}=6$ with the super-renormalization condition, 
assuming the contribution of pure gravity is dominant and tuning level is weak. 
It can avoid current experiments covering the inverse square law of 
Newtonian gravity. If future experiments can find modified gravity 
in short range, it can be a clue of ghost partner as well as extra dimension. 
A difference from the models with extra dimension is that it can predict 
a modification for the gravity on the scale of galaxy group and cluster. 

The scheme with ghost partner is indeed powerful, 
but so far, it is not used widely as the unitarity violation is crucial for physics. 
We have resolved a main difficulty of the problem 
and it can be a good approach to the hierarchy problem. 
By choosing proper physical state with setting the relation with wave packets, 
positive-semi definite norm and independence on the $S$-matrix 
are realized. As we did not provide a strict proof of the unitarity, 
there might be some faults. In particular, it is curious that 
zero-norm states indirectly affect for physics by giving the upper bound of energy. 
This is not seen in other quantum theories and possibly meets a problem. 
All the things we have done in this paper are weak statement and prediction of 
possible new physics toward quantum gravity in low energy.

\end{document}